\documentclass[useAMS,usenatbib]{mn2e}

\usepackage{graphicx}
\usepackage{pdflscape}
\usepackage{longtable}
\usepackage{amssymb}
\usepackage{subfloat}

\title{Sensitive CO(1-0) Survey in Pegasus-Pisces Reduces CO-Dark Gas Inventory by Factor of Two}
\author[Donate and Magnani]
{Emmanuel Donate$^{1}$ \thanks{E-mail: edonate@uga.edu (ED)}  and Loris Magnani$^{1}$ \\
$^{1}$Department of Physics and Astronomy, University of Georgia, Athens, GA 30602, USA}
\begin{document}


\pagerange{\pageref{firstpage}--\pageref{lastpage}} \pubyear{2002}

\maketitle

\label{firstpage}

\begin{abstract}

We conducted high-sensitivity, high-velocity resolution CO(1-0) observations in a region containing a portion of the diffuse molecular  cloud MBM 53 to determine
whether weak CO emission was present.  The results of our observations increase the amount of CO-detectable molecular gas in the region by a factor 
of two.  The increased molecular mass for the cloud, if applicable
to the molecular clouds in the entire Pegasus-Pisces region, decreases the dark molecular gas content from 58\% of the total H$_2$ mass to $\sim$ 30\%.
If the results for MBM53 are applicable to other diffuse clouds, then the fraction of dark gas directly detectable via sensitive CO(1-0) observations in diffuse 
molecular clouds is similar to that predicted by models for Giant Molecular Clouds.

\end{abstract}



\section{Introduction}
The idea that substantial molecular gas is present in the interstellar medium (ISM) but is not detectable
by the CO(1-0) emission line has become fairly prevalent in the last decade.   This component has 
become known as ``dark gas", a term first suggested in a paper describing its properties and extent
by Grenier, Casandjian and Terrier (2005; hereafter GCT)\footnote{Some of the dark gas may be in 
atomic form and not readily detectable by the 21 cm line because of opacity effects.  In this paper we will
focus on the {\it molecular} dark gas.}.  
GCT  used a combination of gamma-ray and infrared data to identify regions with cold dust.  By comparing these with
maps of CO and H{\small {\rm I}} emission, they found that a significant fraction of the H$_2$ mass is 
not traced by the CO(1-0) observations.  Their main conclusion is that the dark gas mass in the Milky Way
is comparable to the molecular mass detected by CO(1-0) emission.  More recent studies corroborate this
(Abdo et al. 2010; Planck Collaboration, Planck Early Results, 2011, Paper XIX).

This phenomenon is predicted by
models of photo-dissociation regions (PDRs) (e.g., Hollenbach \& Tielens 1997).
The depths into a PDR at which H$_2$ and CO form depend directly on the intensity of the interstellar 
radiation field and the volume density of hydrogen nucleons.  It is convenient from the observational 
point of view to talk about depth into a PDR in terms of extinction or color excess.  
Liszt (2014) ascribes
a break in slope in the N(H{\small {\rm I}}) - E(B-V) relation at E(B-V)$\sim$ 0.1 mag to the onset of H$_2$ formation.
For a standard interstellar radiation field (e.g., Draine 1978), the models of van Dishoeck \& Black (1988) 
show a significant rise in both N(CO) and the CO/H$_2$ ratio at N(H$_2$) $\sim$ 1 x 10$^{21}$ cm$^{-2}$,
equivalent to E(B-V) $\approx$ 0.3 mag or A$_V$ $\approx$ 1 mag.     
Thus, under normal conditions, CO-dark gas should be found in regions with 0.1 $\lesssim$ E(B-V) $\lesssim$ 0.3 mag.

van Dishoeck 
\& Black defined {\it diffuse} molecular clouds as those with A$_V \le$ 1 mag. 
Before the development of sensitive millimeter-wave receivers, observations of diffuse molecular clouds
were confined to optical and UV absorption lines.
However, under certain conditions, CO(1-0) emission can trace  diffuse molecular gas fairly easily 
(e.g., Liszt, Pety, and Lucas  2010), complicating the picture described above. But even by
the late 1980s, CO(1-0) detections in regions with A$_V < 1$ mag were possible.  For instance,
many of the high-latitude clouds identified by Magnani, Blitz, and Mundy (1985) have A$_V <$ 1 mag
and yet are readily detectable by CO(1-0) emission.  More recently, Chastain et al. (2006), Liszt and Pety (2012),
and Cotten and Magnani (2013) detected
CO(1-0) lines from regions with E(B-V) as low as  0.1 mag.   Clearly, under certain conditions to be explained, 
CO(1-0) emission can trace gas in the dark molecular regime.

One of the fields studied by GCT  includes 1888 square degrees centered on Pegasus-Pisces for which 
they obtain the ratios of CO-detectable molecular gas, to CO-dark gas, to atomic gas.  We will abbreviate these ratios as
H$_2$:DG:HI .  The CO-detectable
molecular gas was determined from estimates of the masses of some of the mapped high-latitude 
clouds in the region, and GCT derive H$_2$:DG:HI  to be 0.7: 1.0: 3.4, with a total
mass of 1.6 $\times$ 10$^4$ M$_\odot$.   
However, of the many molecular clouds in the region, GCT only included the MBM 53-55 complex and its
environs as mapped by Yamamoto et al. (2003 - hereafter Y2003; 2006).  Blair et al. (2017) examine the Magnani et al. (2000) high-latitude
CO(1-0) survey and, by including  all known molecular clouds in the region, revise the H$_2$:DG:HI ratios to 0.9: 1.0: 3.8.
Even with this revision, the CO-dark gas fraction is still significant.
But are the dark gas regions described by GCT really undetectable in CO(1-0)?
A key factor is the sensitivity of the CO(1-0) observations. 
While the total mass in hydrogen nucleons for Pegasus-Pisces derived by GCT may be correct, the breakdown 
into H$_2$:DG:HI depends on the sensitivity of the CO observations of the molecular clouds in the region.
If more sensitive observations of these clouds were available, the CO-detectable molecular mass
 might increase at the expense of the CO-dark gas fraction.

In this paper we examine whether sensitive CO(1-0) observations can reveal the presence of at least some of the dark gas in 
a high-latitude region containing a portion of a diffuse molecular cloud (MBM 53).   
In section 2, we discuss the Magnani et al. (2000)
Southern Galactic Hemisphere high-latitude CO(1-0) survey (SGH-HL) and the region in Pegasus-Pisces that overlaps one of the
GCT fields. We describe the observational set-up in section 3 and, in section 4, we compare our results with those from GCT.
We then estimate the fraction of dark gas detectable by very sensitive CO(1-0) observations.  A discussion and summary is presented in section 5.


\section{The SGH high-latitude survey in Pegasus-Pisces}

GCT derived their quantitative conclusions on the dark gas by examining seven nearby areas, one of which 
substantially overlaps a portion of the SGH-HL survey (see Figure 1). The area called Pegasus by GCT  and Pegasus-Pisces
in this paper is bounded by the black box.  
Of the 4982 grid points observed in the SGH by Magnani et al. (2000), 2080 points are located within the region that overlaps
84\% of the GCT Pegasus field\footnote {The GCT Pegasus region extends over $-60^\circ \le b \le -26^\circ$, but the SGH-HL CO 
survey begins at $b = -30^\circ$.}.  In that region, Magnani et al. (2000) recorded 48 CO detections with an rms in T$_{mb}$ of 0.1 K  for a detection rate of 0.023.  
 
 GCT quote values for the molecular mass in the 
 Pegasus-Pisces region in CO-detectable H$_2$ and CO-dark H$_2$ of 2200 M$_\odot$ and 3100 M$_\odot$, respectively.   
 Their estimate of the CO-detectable H$_2$ in the region comes solely from the MBM 53-55 cloud complex as mapped
 by Y2003.  We presume that the difference in the molecular mass estimate of the complex between GCT  and the
 Y2003 value is due to GCT using a value of X$_{CO}$\footnote{   X$_{CO}$ is the empirically derived conversion factor between N(H$_2$) and
 the velocity-integrated CO(1-0) main beam antenna temperature.} =  1.74 $\times$ 10$^{20}$ cm$^{-2}$ [K km s$^{-1}$]$^{-1}$ rather than
 the value of 1.0 $\times$ 10$^{20}$ cm$^{-2}$ [K km s$^{-1}$]$^{-1}$ used by Y2003.
 Despite the presence of other clouds in the region (see Blair et al. 2017), we assume, for the moment, that the total mass in hydrogen nucleons 
 detected by GCT is accurate,  hence any increase in the CO-detected gas in the MBM 53-55 clouds must come at the expense of the dark gas mass.  
 
 As mentioned above, the key factor in  determining the amount of CO-detected gas is the sensitivity of the CO(1-0) observations.  More sensitive (lower
 rms) CO observations should yield new detections of molecular gas at cloud edges.
 For example, the high-latitude cloud MBM 40 was sampled over 103 lines of sight in the CO(1-0) line
at sensitivities of  0.05 - 0.11 K (1$\sigma$ rms in  T$_R^*$, or 0.06 - 0.13 K in T$_{mb}$) by Cotten \& Magnani (2013). 
Sampling was over the whole cloud, including the 
outermost parts of the cloud with E(B-V) $\lesssim$ 0.12 mag.  There, weak CO emission was present in 13 out of 21 lines of sight.
The antenna temperature, T$_R^*$, ranged from 0.11 - 2.29 K for the 13 detections with 7 of those
ranging from 0.11 - 0.61 K.  Many of these latter detections would have been missed by conventional mapping which typically
has rms antenna temperature of a few tenths of a K.  By including the molecular mass detected by the sensitive CO observations of the outermost region
of MBM 40, the cloud mass estimate increased from 20 M$_\odot$ (Chastain 2007) to 32 M$_\odot$.  

To determine how much molecular gas could be detected by a more sensitive CO survey at high Galactic
latitude, we re-observed a small region of the
SGH-HL survey in
Pegasus-Pisces.  The SGH-HL survey has 2080 observed lines of sight
within the black box in Figure 1 with T$_{mb} \approx$ 0.1 K for the 1$\sigma$ rms value.  Since our goal was to improve this by at least
a factor of 4-5, we had to choose a significantly smaller region to sample.  We wanted an area containing at least a portion of a 
known molecular cloud (because GCT
emphasized that dark gas regions 
``...surround all the nearby CO clouds and bridge the dense cores
to broader atomic clouds...").
Moreover, the only cloud complex recognized by GCT in the Pegasus-Pisces region in Figure 1 is MBM 53-55, thus, 
by choosing a region containing at least part of this complex, we can
directly compare how our mass estimates for a portion of this complex change with more sensitive observations.  
We settled on the region bounded by $80.3^\circ \le \ell \le99.0^\circ$; $-34^\circ \le b \le -30.0^\circ$.  This area includes
five detections of MBM 53 from the SGH-HL survey.   
The location of the 88 points we observed is shown in Figure 2.  The five CO(1-0) detections from the SGH-HL survey 
arising from the diffuse high-latitude cloud MBM 53 are noted in the figure.


\section{Observations}

The CO(1-0) observations were made at the 12 m mm-wave telescope of the Arizona Radio Observatory on Kitt Peak.
We used the 3 mm  ALMA Type Band 3 dual-polarization receiver with system temperatures on the sky typically less
than 200 K (single sideband).  
The antenna temperature scale, T$_A^*$ (see Kutner and Ulich 1981),  is set by the chopper wheel method and
is given at the telescope as T$_R^*$, the antenna temperature corrected for  spillover and scattering.  Conversion to
the main beam brightness temperature, T$_{mb}$ is via T$_R^* / \eta_{mb}$, where $\eta_{mb}$ is the main beam efficiency (at 115 GHz, 
$\eta_{mb} \approx 0.85$).
We assume that the source fills the beam so that the beam filling factor is unity.  The backend spectrometers were the Millimeter 
Autocorrelator (MAC) and the 250 kHz filter banks.  The MAC was configured to have 200 MHz total bandwidth (150 MHz of usable
bandwidth) over 6144 channels for a frequency resolution of 24.4 kHz per channel equivalent to 0.063 km s$^{-1}$.
The 250 kHz filter banks were used as a backup for the MAC and have a velocity resolution per channel of 0.65 km s$^{-1}$.
A fiducial position ($\ell = 92.4^\circ; b = -34.0^\circ$) with a strong CO line was monitored each day and a Gaussian fit to the
line showed variations in peak T$_R^*$ of less than 5\% in clear weather.

The observations were made in on-off mode with an absolute off-position with the lowest reddening in the region  [E(B-V) = 0.03 mag
at (R.A., Dec) = (22$^h$ 37$^m$ 37.6$^s$; 23$^\circ$ 35$^\prime$ 20.2${^{\prime\prime}}$)] as
given by the Schlegel, Finkbeiner, and Davis (1999) dust maps. At 115 GHz, the beam size of the 12-m telescope is 55$^{\prime\prime}$.
The SGH-HL survey was sampled every degree in Galactic longitude and latitude at a resolution of 8.4$^\prime$
for an under sampling of a factor of $\sim$60.  A meaningful comparison between the CO observations from the
two telescopes requires that we ``map" the 8.4$^\prime$ beam with our ARO 12-m telescope observations.   We thus sampled each of the 88 points in
Figure 2 using the thirteen-point pattern shown in Figure 3.   Although the areal ratio is approximately
0.155, the thirteen point sampling pattern was the best compromise between  covering the 8.4$^\prime$ beam from the
SGH-HL survey and achieving the factor of 5 improvement in sensitivity we desired.
There could be even more CO emission in the region if isolated, small-scale structures exist in the periphery of the cloud.  This issue
will be addressed in a separate paper.

A comparison of the integrated
antenna temperature for the five detected positions from SGH-HL with our thirteen-point pattern averaged to form a single spectrum for
each position shows very good agreement (see Table 1 and discussion in \S 3.1).   With two minutes on source and two minutes off, each
summed spectrum consisted of 26 minutes on-source resulting in rms values per channel for the MAC of 11 - 45 mK.

The results of our observations are shown in Table 2.  Fifteen new detections
were made for a total of 20 out of 88 observed lines of sight detectable in the CO(1-0) line.  Figure 2 shows the distribution of the
new and previously known detections.   Representative spectra of some of the new detections are shown in Figure 4{\it a-c}.

\begin{table*}
\centering
\begin{minipage}{140mm}
 \caption{W$_{CO}$ for the five positions in common between the SGH-HL and the current survey.}
 \begin{tabular}{@{}lcccc@{}}
 \hline
	Position		&     	$\ell$	&	$b$		&  W$_{CO}^a$(SGH-HL) 	& W$_{CO}^a$(this paper) 	 \\
				&	deg		&	deg		& K km s$^{-1}$		& K km s$^{-1}$		 \\
G92.4$-$34		&	92.4		&	$-$34.0	& 	4.09$\pm$0.39		&	4.41$\pm$0.19		 \\
G93.5$-$32		&	93.5		&	$-$32.0 	&	0.57$\pm$0.39		&	0.50$\pm$0.06		 \\
G94.6$-$34		&	94.6		&	$-$34.0	&	1.44$\pm$0.40		&	1.49$\pm$0.24	 	 \\
G96.8$-$30		&	96.8		&	$-$30.0	&	0.46$\pm$0.50 	&	0.86$\pm$0.10		 \\
G99.0$-$31		&	99.0		&	$-$31.0	&	1.00$\pm$0.39		&	1.27$\pm$0.18		 \\
				&			&			&					&					 \\
Total				&  ---			& ---			&	7.56$\pm$0.93		&	8.53$\pm$0.37		\\
\hline
$^a$ W$_{CO}$ is defined as $\int T_{mb} dv$
\end{tabular}
\end{minipage}	
\end{table*}

\begin{table*}
\centering
\begin{minipage}{140mm}
 \caption{Gaussian-fit parameters and W$_{CO}$ for CO(1-0) detections in our survey.}
 \begin{tabular}{@{}lcccllll@{}}
	Position		&     	$\ell$	&	$b$   	&   T$_R^*$  	&   $\Delta$v	&     v$_{LSR}$  &  W$_{CO}^a$ 	& Notes  	\\
				&	deg	&	deg		&	K		&	km s$^{-1}$	&	km s$^{-`}$	& K km s$^{-1}$ \\
\hline
G92.4$-$30		&	92.4	&	$-$30.0	&	0.14$\pm$0.04		&	1.85		&	$-$9.99	&	0.33$\pm$0.13		&	redshifted and blueshifted wings on profile$^c$ 	\\
G93.5$-$30		&	93.5	&	$-$30.0	&	0.36$\pm$0.04		&	2.12		&	$-$10.09	&	0.96$\pm$0.15		&										\\
G96.8$-$30		&	96.8	&	$-$30.0	&	0.62$\pm$0.05		&	1.10		&	$-$14.90	&	0.86$\pm$0.10		&										\\
				&		&			&	0.35$\pm$0.09$^b$	&	1.25		&	$-$14.66	&	0.46$\pm$0.50		& 	Magnani et al. (2000)					\\
G92.4$-$31		&	92.4	&	$-$31.0	&	0.25$\pm$0.03		&	1.11		&	$-$5.44	&	0.35$\pm$0.06		&	first component$^d$ 						\\
				&		&			&	0.14$\pm$0.03		&	0.73		&	$-$8.32	&	0.13$\pm$0.04		&	second component						\\	
G93.5$-$31		&	93.5	&	$-$31.0	&	0.65$\pm$0.05		&	0.92		&	$-$3.15	&	0.75$\pm$0.08		&	first component   						\\
				&		&			&	0.43$\pm$0.05		&	0.95		&	$-$4.72	&	0.51$\pm$0.08		&	second component 						\\
G97.9$-$31		&	97.9	&	$-$31.0	&	0.097$\pm$0.033	& 	0.88 		&	$-$11.61	&	0.11$\pm$0.05		&										\\
G99.0$-$31		&	99.0	&	$-$31.0	&	0.49$\pm$0.05		&	2.06		&	$-$8.18	&	1.27$\pm$0.18		&	blueshifted wing on profile 				\\
				&		&			&	0.27$\pm$0.07$^b$	&	3.41		&	$-$8.47	&	1.00$\pm$0.39		& 	Magnani et al. (2000)   		 			\\
G92.4$-$32		&	92.4	&	$-$32.0	&	0.23$\pm$0.03		&	1.01		&	$-$2.33	&	0.29$\pm$0.05		&	first component - redshifted wing on profile	\\
				&		&			&	0.11$\pm$0.03		&	0.91		&	$-$7.86	&	0.13$\pm$0.05		&	second component 						\\
G93.5$-$32		&	93.5	&	$-$32.0 	&	0.43$\pm$0.03		&	0.63 		&	$-$4.67	& 	0.34$\pm$0.03 	& 	 first component						\\
				&		&			&	0.14$\pm$0.03		&	0.89		&	$-$3.95	&	0.16$\pm$0.05		&	second component 						\\
				&		&			&	0.37$\pm$0.07$^b$	&	1.46		&	$-$4.60	&	0.57$\pm$0.39		& 	Magnani et al. (2000)					\\
G94.6$-$32		&	94.6	&	$-$32.0 	&	0.48$\pm$0.03		&	0.71		&	$-$2.91	& 	0.43$\pm$0.04 	&										\\
G95.7$-$32		&	95.7	&	$-$32.0	&	0.32$\pm$0.03		&	2.86		&	$-$6.40	&	1.15$\pm$0.15		&										\\ 
G99.0$-$32		&	99.0	&	$-$32.0	&	0.31$\pm$0.03		&	1.13		&	$-$6.71	&	0.44$\pm$0.06		&										\\
G90.2$-$33		&	90.2	&	$-$33.0	&	0.38$\pm$0.03		&	0.60		&	$-$0.56	&	0.28$\pm$0.03		&										\\
G92.4$-$33		&	92.4	&	$-$33.0	&	0.076$\pm$0.028 	&	2.05		&	$-$5.22	&	0.20$\pm$0.10		&	first component							\\
				&		&			&	0.077$\pm$0.028 	&	1.20		&	$-$8.63	&	0.12$\pm$0.06		&	second component						\\
G97.9$-$33		&	97.9	&	$-$33.0	&	0.31$\pm$0.04		&	0.98		&	$-$1.62	&	0.38$\pm$0.07		&	first component							\\
				&		&			&	0.10$\pm$0.04		&	1.36		&	$-$3.58	&	0.17$\pm$0.09		&	second component						\\
G92.4$-$34		&	92.4	&	$-$34.0	& 	1.52$\pm$0.05		&	2.01		&	$-$7.48	&	3.85$\pm$0.18		&	first component							\\
				&		&			&	0.54$\pm$0.05		&	0.83		&	$-$7.27	&	0.56$\pm$0.07		&	second component						\\
				&		&			&	1.34$\pm$0.07$^b$	&	2.84		&	$-$7.47	&	4.09$\pm$0.39		&	Magnani et al. (2000)					\\
G93.5$-$34		&	93.5	&	$-$34.0	&	0.61$\pm$0.03		&	1.46		&	$-$4.21 	&	1.12$\pm$0.08		&	blueshifted wing centered at $\sim -$7 km s$^{-1}$ \\
G94.6$-$34		&	94.6	&	$-$34.0	&	0.51$\pm$0.06		&	1.81		&	$-$6.06	&	1.17$\pm$0.19		&	first component							\\
				&		&			&	0.20$\pm$0.06		&	1.27 		&	$-$2.92	&	0.32$\pm$0.14		&	second component						\\
				&		&			&	0.49$\pm$0.07$^b$	&	2.77		&	$-$6.13	&	1.44$\pm$0.40		&	Magnani et al. (2000)					\\
G95.7$-$34		&	95.7	&	$-$34.0	&	0.31$\pm$0.04		&	0.69		&	$-$4.89	&	0.27$\pm$0.05		&										\\
G96.8$-$34		&	96.8	&	$-$34.0	&	0.37$\pm$0.04		&	2.88		&	$-$12.58	&	1.34$\pm$0.21		&	first component							 \\
				&		&			&	0.21$\pm$0.04		&	1.66		&	$-$7.07	&	0.44$\pm$0.12		& 	second component						 \\
				&		&			&					&			&			&					&	redshifted wing centered at $\sim -$4.5 km s$^{-1}$ \\
\hline
				&		&			&					&			&			&					&	     									\\
\end{tabular}
$^a$ W$_{CO}$ is defined as $\int T_{mb} dv$

$^b$ T$_{mb}$ from Magnani et al. (2000)

$^c$ Some of the line profiles were not well-fit by one or two Gaussians and showed weak emission in either the blue or red (or both) wings
of the profile.  In these instances, the quoted value of W$_{CO}$ is an underestimate since it is estimated based on the Gaussian fits.

$^d$ Some of the line profiles showed two distinct components.
\end{minipage}	
\end{table*}

\subsection{Comparison with earlier results}

Five of the ARO detections had been previously detected with the Harvard-Smithsonian Center for Astrophysics
1.2 m telescope (Magnani et al. 2000).  As can be seen from Table 1, the agreement between the two sets of observations is quite good.  The SGH-HL
observations were about a factor of five less sensitive in T$_{mb}$ while the difference in W$_{CO}$  1$\sigma$ uncertainties improves for the ARO observations
by factors of about 2 to 6.  For the ARO data, W$_{CO}$ is calculated from the parameters of the Gaussian fit as 1.07[T$_{mb}\Delta$v(FWHM)].The total W$_{CO}$ for the  five SGH-HL detections is 7.56 $\pm$ 0.93 K km s$^{-1}$ and, for the ARO observations, 8.53 $\pm$ 
0.37 K km s$^{-1}$, so that estimates of the mass based on these results would be consistent.  However, because of the limited sampling of the SGH-HL beam by 
our current observations we note that the ARO results are lower limits.

When looking at Table 2, some of the new ARO detections have W$_{CO}$ values that should have been detected by the SGH-HL survey.
An example of this is position G902-33, with T$_R^* \approx$ 0.4 K, equivalent to T$_{mb} \approx$ 0.5 K.  The rms sensitivity of the SGH-HL 
survey was 0.1 K and should have resulted in a detection.  However, the SGH-HL spectrum shown in top half of Figure 5 shows no evidence of the line.  This issue can be 
understood by noting the narrowness of the line at this position (0.6 km s$^{-1}$) compared to the velocity resolution of the filter banks used in the SGH-HL
survey: 0.65 km s$^{-1}$.  The line, if it had been detected would have been in only 1-2 channels, making it more susceptible to noise fluctuations than if
higher frequency resolution was employed.  Figure 5 shows the non-detection from SGH-HL compared to the spectrum from the data presented in this
paper.  The surprise here is the importance of observing diffuse molecular cloud at high velocity resolution.

\subsection{The mass of MBM 53-55}

Since we are re-observing a part of the MBM 53-55 region sampled in CO(1-0) by Magnani et al. (2000), any new detections will increase the mass of that
cloud complex.  Here we determine the mass of the complex from the original Magnani et al. (2000) data and compare our result to that of Y2003.
The Pegasus-Pisces region (with northernmost boundary $b = -30^\circ$) shown in Figure 1 covers 0.5749 steradians.  The SGH-HL survey covers this region with 2080 points.
Of the 48 detections in this region, 23 are associated
with the MBM 53-55 cloud complex (see Magnani et al. 2000) providing an estimate of the solid angle of the cloud.  We can derive the mass of these 
clouds using the average W$_{CO}$ value
from SGH-HL (2.47 K km s$^{-1}$), the distance to the clouds (150 pc), a value for the mean molecular weight of 2.8, and X$_{CO}$ = 1.0 $\times$ 10$^{20}$ cm$^{-2}$ [K km s$^{-1}$]$^{-1}$.
The latter three values were chosen to be identical to those used by Y2003. In this fashion we obtain a mass of 790 M$_\odot$
compared to 1200 M$_\odot$ by Y2003. In comparing these results, we note that the region surveyed by Y2003 is slightly larger 
than ours, extending to $b = -29^\circ$, but the clouds there are not very prominent and contribute little mass to the complex.  The discrepancy
in the mass estimate is also not due to sensitivity considerations since the SGH-HL survey has an rms of 0.1 K versus 0.5 for Y2003.  The reasons are likely
to be either the peculiar sampling method used by Magnani et al. (2000) where the off position for the spectra was chosen to be one degree south of 
the on scan (see discussion by Hartmann, Magnani, and Thaddeus 1998),  beam dilution of small structures at the cloud edges, and/or non-detections
of narrow lines because of the relatively poor velocity resolution of the SGH-HL survey (see discussion above).  The SGH-HL survey
has a beam of 8.4$^\prime$ and a velocity resolution of 0.65 km s$^{-1}$  versus 2.6$^\prime$ and 0.1 km s$^{-1}$, respectively, for Y2003.   

GCT increase the Y2003 mass estimate to 2200 M$_\odot$ by using a larger value of X$_{CO}$.   If we  increase the SGH-HL result by the same factor,
the mass of the clouds would then be 1450 M$_\odot$.   This is the result we will use for the CO-detected mass in the MBM 53-55 complex based
on  the SGH-HL survey.


\section{Results and Implications}

The 15 new detections from our ARO observations more than double the W$_{CO}$ value from the SGH-HL survey (from 8.53 K km s$^{-1}$ to 
18.42 K km s$^{-1}$, see Tables 2 and 3).  Since the cloud mass is directly proportional
to W$_{CO}$, if our results are typical for the rest of the cloud complex, then the overall mass would increase by a factor of 2.16.   Given the results of \S 3.2, this would mean that the
mass of the MBM 53-55 complex would increase on the basis of our sensitive, high velocity resolution CO observations from 1450 M$_\odot$ to 3130 M$_\odot$.

GCT derive  H$_2$:DG:HI for the
entire Pegasus-Pisces region to be 0.7: 1.0: 3.4.  If the total
mass  from the gamma-ray data is 1.6 $\times$ 10$^4$ M$_\odot$, then the ratio H$_2$:DG:HI  in solar masses would be 2200: 3100: 10,700 M$_\odot$. 
Assuming the atomic hydrogen is well-determined (i.e., there are no opacity issues),  {\it the increase of 930 M$_\odot$ in CO-detectable gas (i.e., from GCT's 
value of 2200 M$_\odot$ to our value of 3130 M$_\odot$) must come at the expense of the CO-dark gas.}
This would reduce the CO-dark gas in the region from GCT's estimate of 3100 M$_\odot$ to 2170 M$_\odot$.  The total molecular gas content (i.e., CO-detectable plus 
dark) remains the same, so the CO-dark gas contribution to the total molecular gas  now decreases to about $\sim$ 40\%.


\begin{table*}
\begin{minipage}{0.5\textwidth}
 \caption{Total W$_{CO}$ from the three surveys.}
 \begin{tabular}{@{}ccc@{}}
 \hline
	Survey						&   	 W$_{CO}^a$  	& rms	 \\
								& K km s$^{-1}$         &   K km s$^{-1}$    \\
								&				&        \\
	
	SGHL-HL survey				&  7.56			&     	0.93         \\
	This paper. SGH-HL detections        	&  8.53                         &	0.37		\\
	This paper, new detections		& 9.89			&  	0.57		\\
\hline
$^a$ W$_{CO}$ is defined as $\int T_{mb} dv$
\end{tabular}
\end{minipage}	
\end{table*}

These results are based on sampling 88 of the 2080 points observed in the SGH-HL survey for this region.   
Our observations focus on the 
northernmost portion of the MBM 53-55 complex which is the only molecular gas considered by GCT in the region.   
The locations of our new detections
(see Figure 2) are, for the most part, in the outer regions of the cloud, where low-intensity CO-emitting gas should be located.  Since the region chosen for our
high-sensitivity observations was chosen randomly along the MBM 53-55 cloud, it seems reasonable to assume that this pattern holds throughout the cloud.  A similar
result was obtained for the diffuse molecular cloud, MBM 40, by Cotten \& Magnani (2013) who derived a mass of 32 M$_\odot$ from sensitive
CO(1-0) observations (rms values 0.05 - 0.09 K). Conventional CO maps of this cloud (i.e., with rms values of 0.1 - 0.7 K in antenna temperature)
yielded mass estimates in the 11-20 M$_\odot$ range (Magnani, Hartmann, \& Speck 1996; Chastain 2005).  

If our results are applicable to other regions, then $\sim$ one-third of the CO-dark gas is not ``dark" as far as CO observations are concerned.  
Unfortunately, the long integration times necessary to detect much of the low-level CO emission reported in this paper, likely
preclude any significant mapping of this portion of the dark gas. 

The best spectroscopic tracer of  dark gas is the [CII] line at 158 $\mu$m
since the integration times are more reasonable (e.g., Langer et al. 2010; Langer et al. 2014).  The other spectroscopic candidates for tracing dark gas are the OH 18 cm main lines and, in
particular, the 1667 MHz line.    Several studies [Barriault et al, 2010; Cotten et al. (2012); and Allen et al. (2015)] have indicated that OH observations
may pick up molecular gas not detectable in CO - but these authors compared their results to conventional CO mapping with rms values $\sim$ 0.1 K.
We believe that sensitive CO(1-0) observations like the ones reported here would be as successful in detecting a portion of the dark gas as OH, and with 
comparable integration times.  

Our results are based on the assumption that the low-level emission we detected during our survey can be analyzed
in the same manner as the stronger emission from the SGH-HL survey (e.g., the same X$_{CO}$ applies to both 
components). However, Liszt and Pety (2012) showed that the CO(1-0)
line can be over luminous in diffuse molecular gas.   If the newly-detected CO emission arises from this type of gas, then 
a lower value of X$_{CO}$ would have to be used to determine its contribution to the molecular mass.
However, Liszt and Pety (2012) state that in diffuse molecular clouds gas the lines of sight with W$_{CO} \ge$
1.5 K km s$^{-1}$ are likely to be over luminous.  None of the new detections described in this paper are over that threshold.
Nevertheless, the issue of what X$_{CO}$ value to use when converting CO(1-0) observations to N(H$_2$) has always 
been thorny.
Further observations,
perhaps using the CH line at 3.3 GHz to determine independently N(H$_2$) (e.g., Magnani and Onello 1995), could settle this question.


\section{Discussion}

By making more sensitive, higher velocity resolution CO(1-0) observations of a region containing a portion of the high-latitude molecular cloud complex, MBM 53-55, we
determine that the mass of molecular gas in that region detectable by the CO(1-0) line increases by about a factor of two.  If that result is applicable to the entire 
cloud, then the increase in molecular mass detectable by CO(1-0) observations increases significantly - at the expense of the dark gas estimate
for that region.  The increase in gas mass comes at the periphery of the cloud and consists of low-surface brightness CO emission. 
We stress that in addition to greater sensitivity, observations at high velocity resolution (at least 0.1 km s$^{-1}$ per channel) are necessary to detect
some of the weak CO emission in diffuse and translucent molecular gas.

Our results are reminiscent of those from mapping the Gemini OB1 molecular cloud complex by Carpenter, Snell, and Schloerb (1995) who find that up to 50\% of the CO
{\it intensity} map may be tied up in low surface brightness gas normally surrounding the dense cores of a GMC.  Perhaps we are seeing this phenomenon at
lower intensity levels given that we are observing a diffuse/translucent molecular cloud rather than a GMC.
In a similar vein,
Liszt, Pety, and Lucas (2010) and Goldsmith et al. (2008) have found that a large fraction of CO emission comes from warm (50 - 100 K), low-density
(100 - 500 cm$^{-3}$), weakly-shielded diffuse molecular gas, even though most of the carbon in these regions is in the form of C$^+$.

Low intensity CO emission arising from more diffuse molecular gas than that traced by conventional surveys may be more commonplace in the
Galaxy than is currently thought.  Dame \& Thaddeus (1994) detected a faint, thick molecular disk in the inner Galaxy, with a ${12}$CO/${13}$CO ratio
more similar to that in diffuse and translucent clouds than in GMCs.   A similar result has been found by Pety et al. (2013) in M51.

Our results are confined to a small area in the MBM 53-55 cloud complex, but
there are other known molecular clouds in the region that were ignored by GCT but detected in the SGH-HL survey.  
Blair et al. (2017) estimate that these clouds contribute an additional
300 M$_\odot$ of CO-detectable gas (at typical rms values of 0.1 K).  This further reduces the dark gas inventory in the region from about 40\%
of the total molecular mass to about 35\%.   If we assume that sensitive CO observations would also double the mass of these clouds, then the
dark gas inventory in the Pegasus Pisces region drops to 30\%.   This value is similar to what Wolfire, Hollenbach, \& McKee (2010) determine on the basis of static
PDR models of GMC envelopes.  It would appear from our results, that the CO-dark gas fraction does not significantly increase in diffuse and translucent gas.
We note that dynamical effects in PDR models could alter the dark gas fraction.

The detection of a significant fraction of CO-dark gas by sensitive CO(1-0) observations is important not just as a proof-of-concept.
As mentioned above, spectral line detections also provide velocity information.  Moreover, the systematics in using CO data are 
completely different than those from gamma-ray and infrared observations.  CO(1-0) observations are not subject to variations in the
gas-to-dust ratio, the uniformity of the cosmic ray environment, or their penetration depth.

We are not advocating that sensitive CO(1-0) observations are the best way to trace dark gas.  It is clear that [CII] observations are the best
way of spectroscopically tracing this component (see Langer et al. 2010;  2014).  
However, there have been many claims that as much molecular gas exists in dark as opposed to CO-detectable
form (e.g., GCT;  Abdo, et al. 2010; Planck Collaboration 2011).  In addition, several authors [Barriault et al. (2010); Allen et al. (2012; 2015)], claim the OH
18 cm main lines are better tracers of diffuse molecular gas than the CO(1-0) line. 
Our results show these claims are premature unless the regions in question have been
observed in CO(1-0) both with adequate sensitivity {\it and} sufficient velocity resolution.



\section*{Acknowledgments}

We thank Professor Lucy Ziurys for generously allocating the observing time to complete this project.
The Kitt Peak 12 meter radio telescope is operated by the Arizona Radio Observatory (ARO), Steward Observatory, University of Arizona.
We also thank Professors Hector Arce and Steven Shore for comments which greatly improved this work.  
Finally, we thank the anonymous referee who raised several points which clarified and improved the manuscript.


\bsp

\clearpage

\begin{figure*}
\includegraphics[width =1\textwidth]{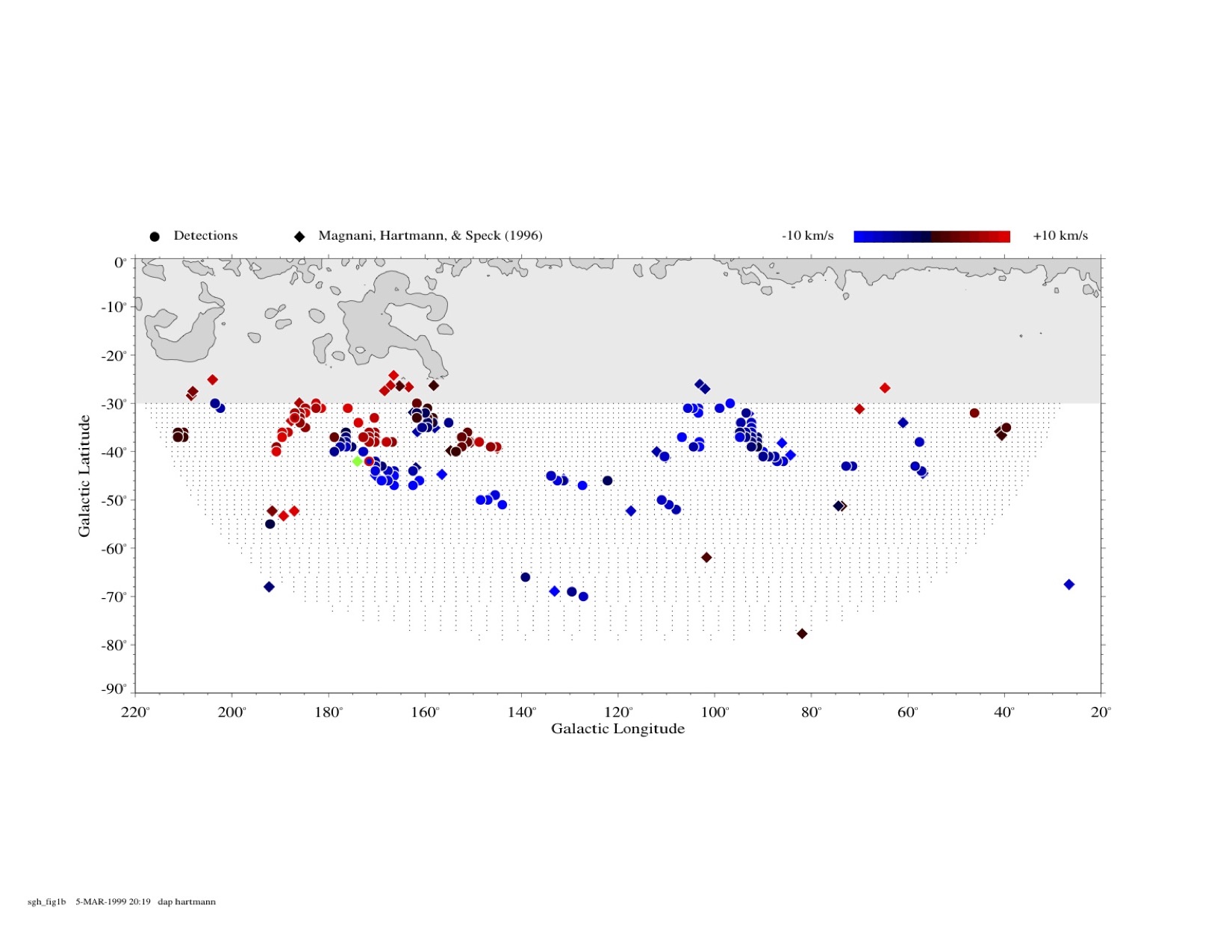}

\caption{The Southern Galactic Hemisphere - High Latitude (SGH-HL) CO(1-0) survey.  The 4982 grid points which were observed by Magnani et al. (2000) are depicted as the
small black dots.  Detections from the survey are the round dots color coded in velocity (see top right corner).  Other detections from Magnani, Hartmann, and Speck (1996) that were
missed by the SGH-HL sampling grid are denoted by diamonds (the green diamond has no velocity information in the literature).   The black box denotes the boundary
of the region in Pegasus-Pisces included in the GCT dark gas calculations (see text).} (A color version of this figure is available in the online journal.)
\label{fig:figure1}
\end{figure*}


\clearpage

\begin{figure*}
\includegraphics[width =1\textwidth]{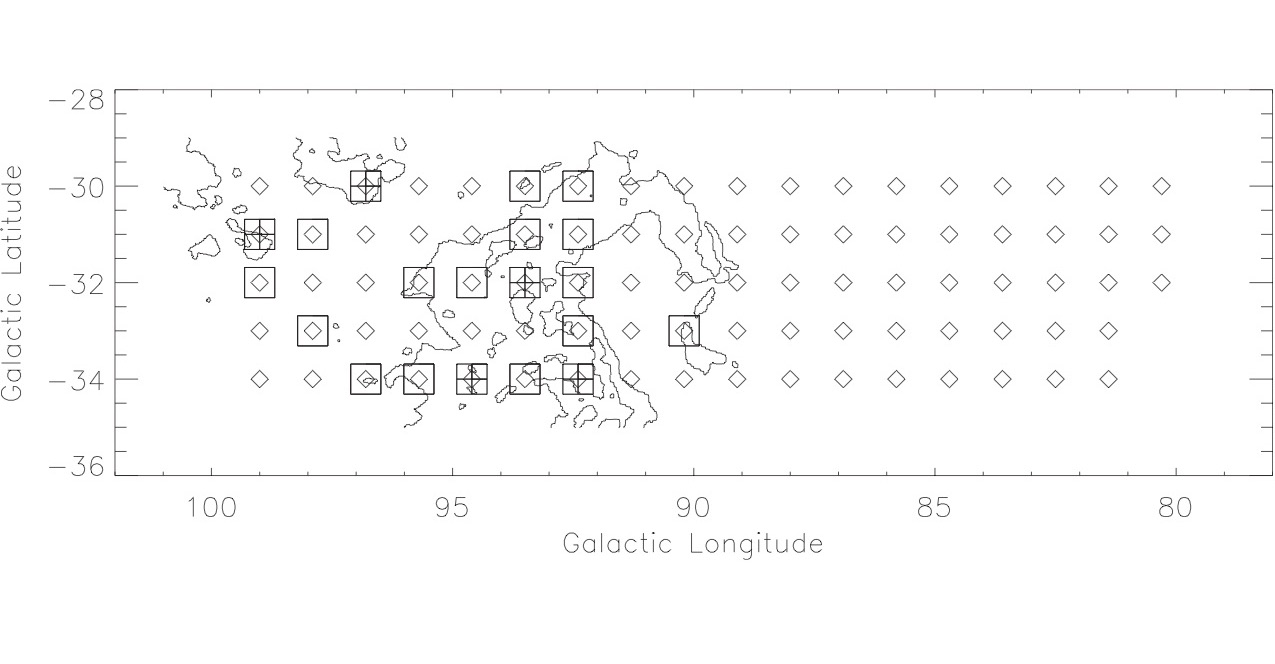}
\caption{Contour map of the northern portion of MBM 53 in E(B-V) from Schlegel, Finkebeiner, and Davis (1999) in linear scale from 0.10 to 0.51 mag in steps of 0.17 mag.  
The diamonds indicate the 88 positions that were observed in CO(1-0) for this paper.  Each position corresponds to an observation from the SGH-HL survey (see \S 2).  
The five CO(1-0) detections from that survey at the T$_{mb}$ level of 0.1 K (listed in Table 1) are denoted  by squares with a plus sign in them. 
The 15 new detections are denoted by squares.  Parameters for all the detections are in Table 2.}
\label{fig:figure2}
\end{figure*}


\clearpage

\begin{figure*}
\center
\includegraphics[width =0.8\textwidth]{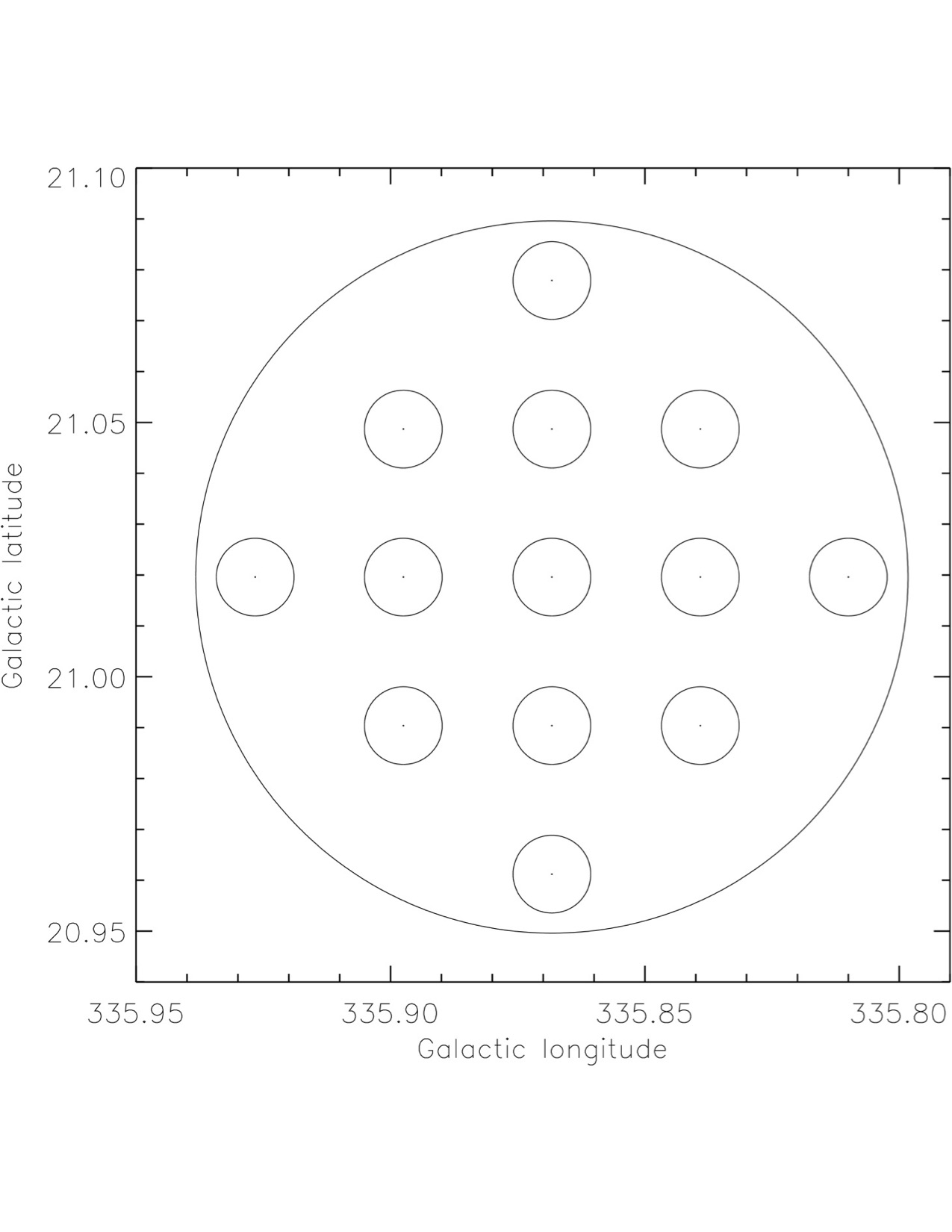}

\caption{Comparison between the beam of the Harvard Smithsonian -- Center for Astrophysics 1.2 m mm-wave telescope (beam size 8.4$^\prime$ at 115 GHz) with the 
13-point beam pattern from the 12-m Arizona Radio Observatory telescope used to sample the larger beam.  The sampling pattern is for one of the 88 observed positions
(see Figure 2) and is representative of how each of those positions were observed in CO, with the final spectrum for that position produced by averaging the 13 individual
spectra.  The beam size of the 12-m at 115 GHz is 55$^{\prime\prime}$.}
\label{fig:figure3}
\end{figure*}


\clearpage

\begin{subfigures}
\begin{figure*}
\includegraphics[width =1\textwidth]{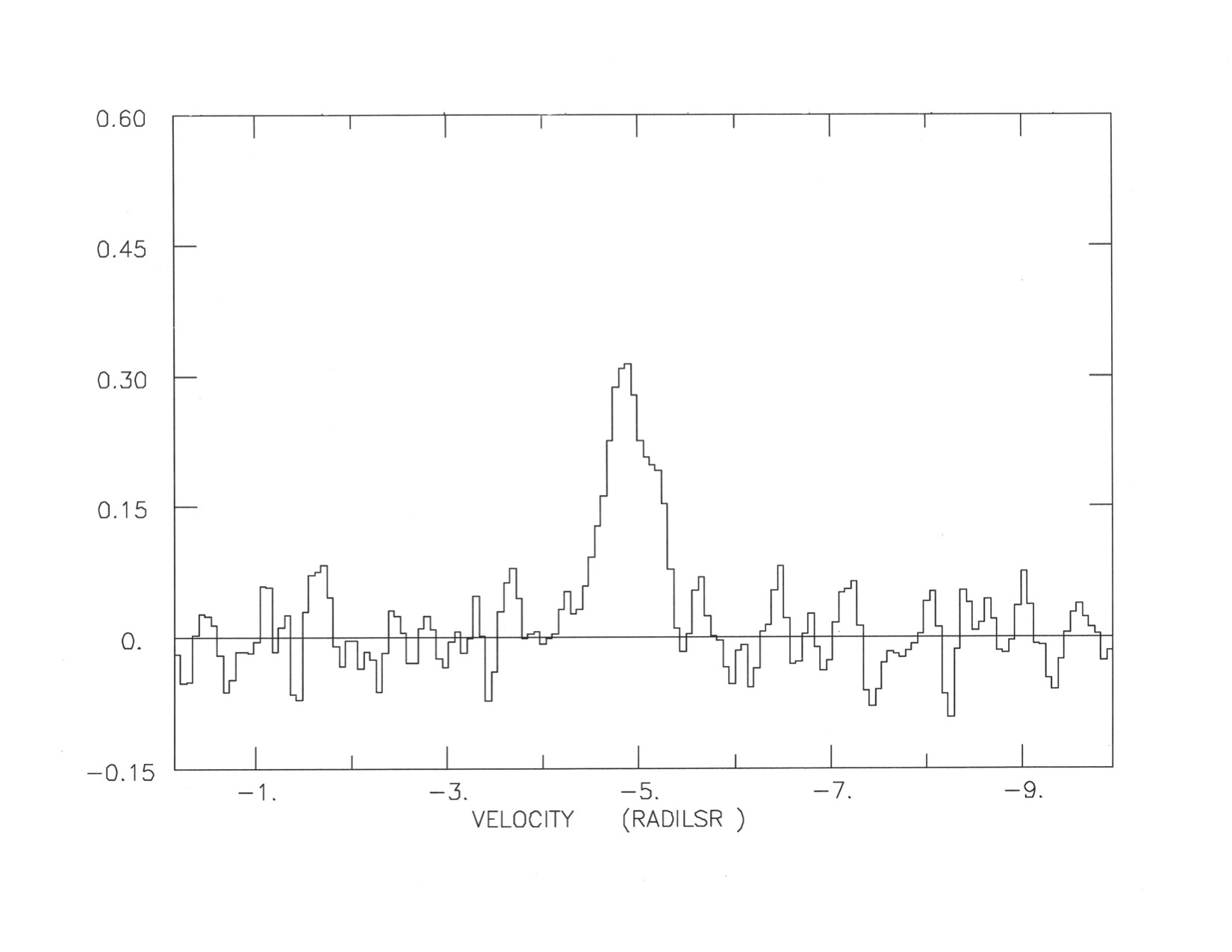}
\caption{CO(1-0) spectrum from position ($\ell, b$) = (95.7$^\circ$, $-$34.0$^\circ$).  The y-axis is antenna temperature in K in the form T$_R^*$
and the x-axis denotes v$_{LSR}$ in km s$^{-1}$.  The velocity resolution and rms per channel are 0.06 km s$^{-1}$ and 39 mK, respectively.}
\label{fig:4a}
\end{figure*}


\clearpage

\begin{figure*}
\includegraphics[width =1\textwidth]{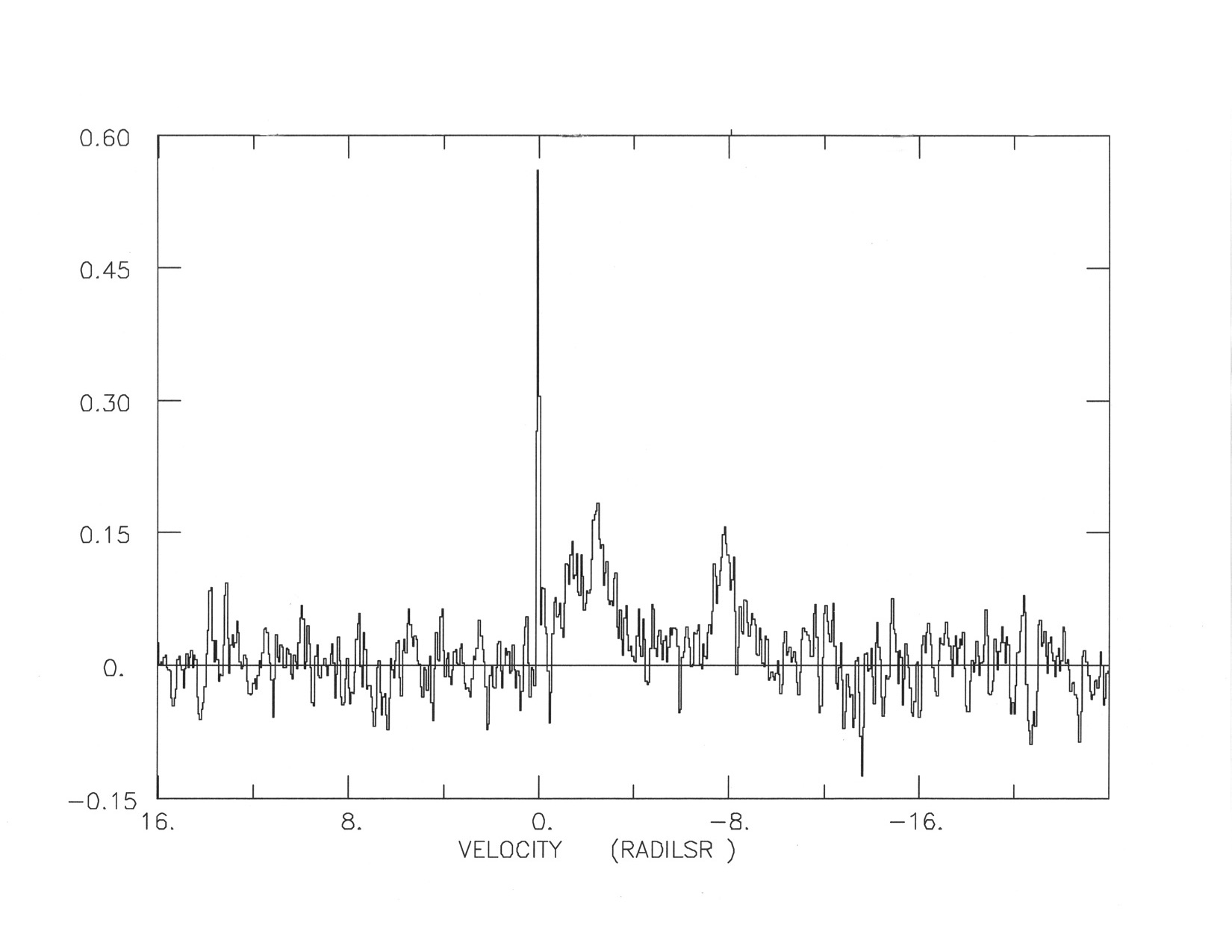}
\caption{CO(1-0) spectrum from position ($\ell, b$) = (92.4$^\circ$, $-$32.0$^\circ$).  The x- and y-axes are as in Figure 5a. The rms is 34 mK per channel.  
The narrow pip at v$_{LSR} \approx$ 0 km s$^{-1}$ is RFI from the system electronics.}
\label{fig:4b}
\end{figure*}


\clearpage

\begin{figure*}
\includegraphics[width =1\textwidth]{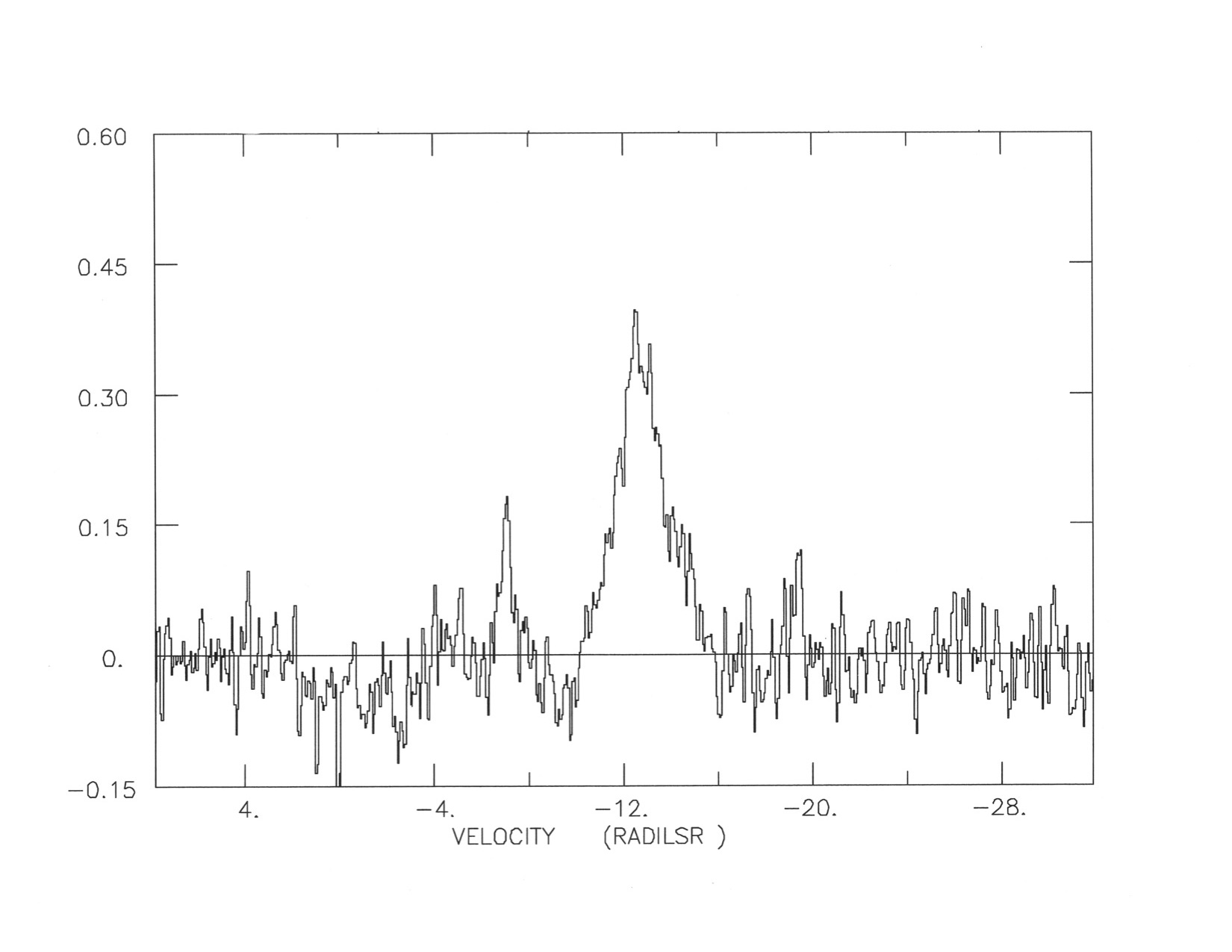}
\caption{CO(1-0) spectrum from position ($\ell, b$) = (92.4$^\circ$, $-$32.0$^\circ$).  The The x- and y-axes are as in Figure 5a. The rms is 37 mK per channel.}
\label{fig:4c}
\end{figure*}
\end{subfigures}

\clearpage

\begin{subfigures}
\begin{figure*}
\includegraphics[width =1\textwidth]{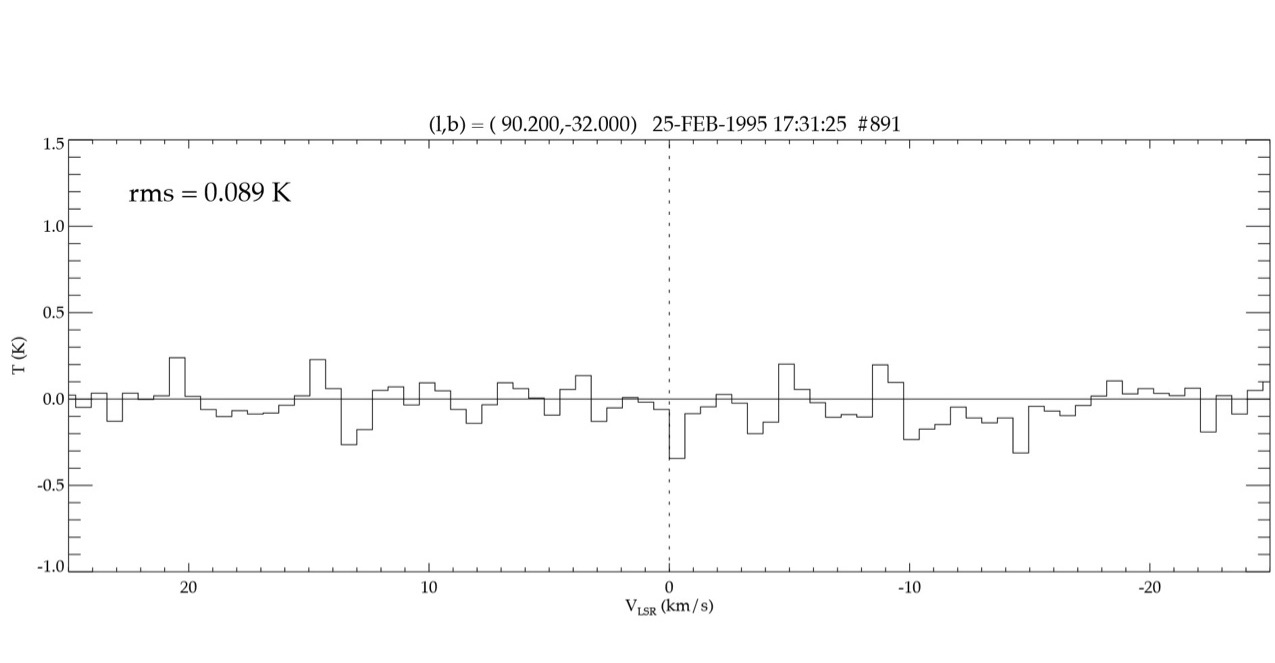}
\caption
{CO(1-0) spectrum from the SGH-HL survey.  The spectrum was made using on-off mode with the on
position at ($\ell, b$) = (90.2$^\circ$, $-$32.0$^\circ$) and the off position at ($\ell, b$) = (90.2$^\circ$, $-$33.0$^\circ$).  
See Hartmann, Magnani, and Thaddeus (1998) for details.  With this observing technique, if a CO(1-0) emission line were to be present in the
off position, it would appear as an absorption line in the spectrum.  The y-axis is antenna temperature in K in the form T$_R^*$.
The velocity resolution is 0.65 km s$^{-1}$.}
\label{fig:figure5a}
\end{figure*}


\clearpage

\begin{figure*}
\includegraphics[width =1\textwidth]{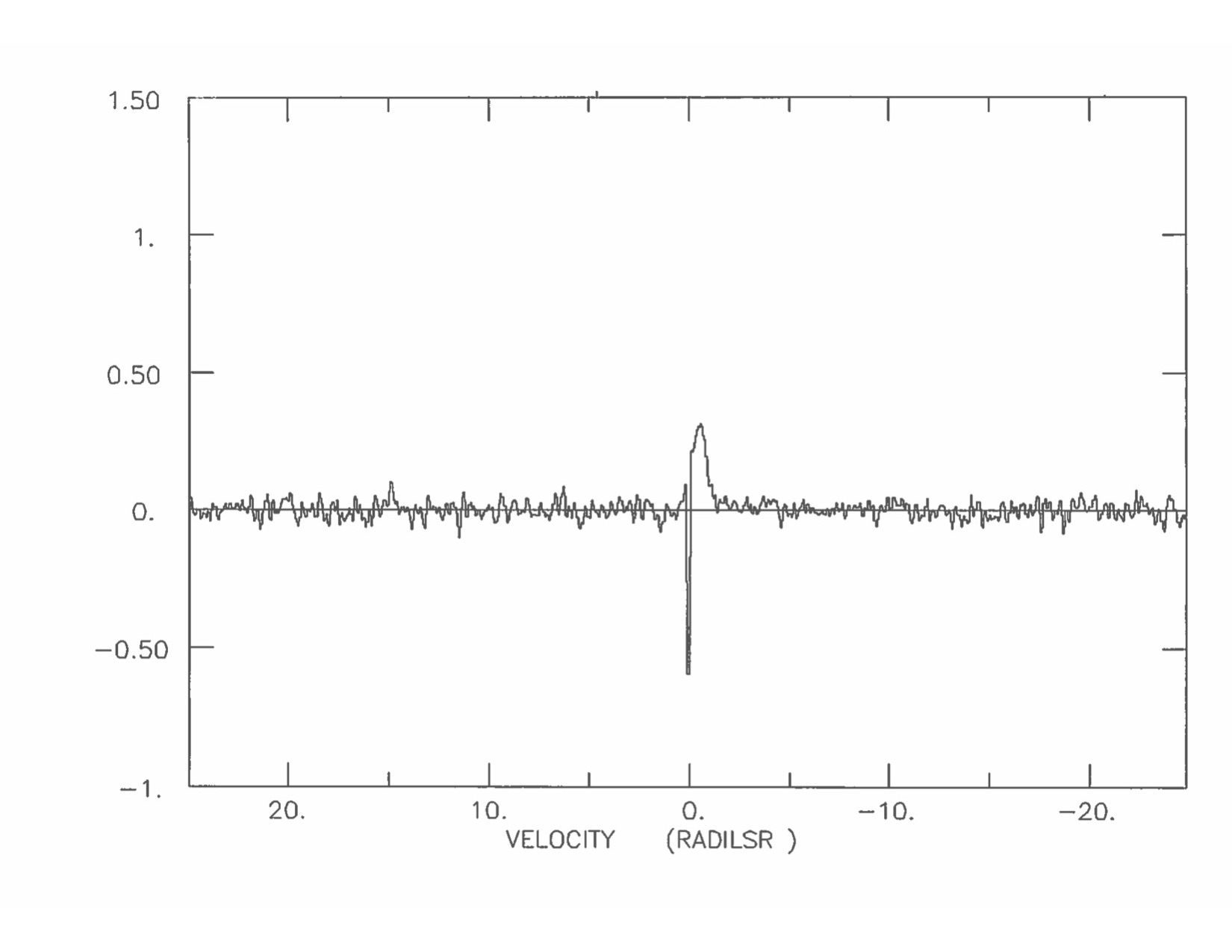}
\caption{CO(1-0) spectrum from position ($\ell, b$) = (92.4$^\circ$, $-$33.0$^\circ$).  The FWHM of the line from Gaussian fitting is 0.60 km s$^{-1}$.
The downward going spike at 0 km s$^{-1}$ is interference.  The y-axis is antenna temperature in K in the form T$_R^*$ and the x-axis is v$_{LSR}$
in km s$^{-1}$. The velocity resolution is 0.06 km s$^{-1}$ and the rms is  25 mK.}
\label{fig:figure5b}
\end{figure*}
\end{subfigures}


\label{lastpage}

\end{document}